\documentclass{PoS}

\usepackage{bm}
\usepackage{mathptmx}
\usepackage{amsmath}
\usepackage{bm}
\usepackage{textcomp}

\DeclareMathAlphabet\mathbfcal{OMS}{cmsy}{b}{n}

\title{Interferometric Radio Measurements of Air Showers with LOPES: Final Results}

\ShortTitle{LOPES: Final Results}

\author{
\speaker{F.G.~Schr\"oder}$^{1}$,
K.~Link$^{1}$,
W.D.~Apel$^{1}$,
J.C.~Arteaga-Vel\'azquez$^{3}$,
L.~B\"ahren$^{4}$,
K.~Bekk$^{1}$,
M.~Bertaina$^{5}$,
P.L.~Biermann$^{5,1}$,
J.~Bl\"umer$^{1,6}$,
H.~Bozdog$^{1}$,
I.M.~Brancus$^{7}$,
E.~Cantoni$^{4,8}$,
A.~Chiavassa$^{5}$,
K.~Daumiller$^{1}$,
V.~de~Souza$^{9}$,
F.~Di~Pierro$^{5}$,
P.~Doll$^{1}$,
R.~Engel$^{1}$,
H.~Falcke$^{10,3,5}$,
B.~Fuchs$^{2}$,
H.~Gemmeke$^{11}$,
C.~Grupen$^{12}$,
A.~Haungs$^{1}$,
D.~Heck$^{1}$,
J.R.~H\"orandel$^{10}$,
A.~Horneffer$^{6}$,
D.~Huber$^{2}$,
T.~Huege$^{1}$,
P.G.~Isar$^{13}$,
K-H.~Kampert$^{14}$,
D.~Kang$^{2}$,
O.~Kr\"omer$^{11}$,
J.~Kuijpers$^{10}$,
P.~{\L}uczak$^{15}$,
M.~Ludwig$^{2}$,,
H.J.~Mathes$^{1}$,
M.~Melissas$^{2}$,
C.~Morello$^{8}$,
J.~Oehlschl\"ager$^{1}$,
N.~Palmieri$^{2}$,
T.~Pierog$^{1}$,
J.~Rautenberg$^{14}$,
H.~Rebel$^{1}$,
M.~Roth$^{1}$,
C.~R\"uhle$^{11}$,
A.~Saftoiu$^{7}$,
H.~Schieler$^{1}$,
A.~Schmidt$^{11}$,
S.~Schoo$^{1}$,
O.~Sima$^{16}$,
G.~Toma$^{7}$,
G.C.~Trinchero$^{8}$,
A.~Weindl$^{1}$,
J.~Wochele$^{1}$,
J.~Zabierowski$^{15}$,
J.A.~Zensus$^{5}$ - 
LOPES Collaboration \\
\llap{$^{1}$} Institut f\"ur Kernphysik, Karlsruher Institut f\"ur Technologie (KIT), Germany\\
\llap{$^{2}$} Institut f\"ur Experimentelle Kernphysik, Karlsruher Institut f\"ur Technologie (KIT), Germany\\
\llap{$^{3}$} Instituto de F\'isica y Matem\'aticas, Universidad Michoacana, Morelia, Mexico\\
\llap{$^{4}$} ASTRON, Dwingeloo, The Netherlands\\
\llap{$^{5}$} Dipartimento di Fisica, Universit\`a degli Studi di Torino, Torino, Italy\\
\llap{$^{6}$} Max-Planck-Institut f\"ur Radioastronomie, Bonn, Germany\\
\llap{$^{7}$} National Institute of Physics and Nuclear Engineering, Bucharest-Magurele, Romania\\
\llap{$^{8}$} Osservatorio Astrofisico di Torino, INAF Torino, Italy\\
\llap{$^{9}$} Universidade S$\tilde{a}$o Paulo, Instituto de F\'{\i}sica de S$\tilde{a}$o Carlos, S$\tilde{a}$o Carlos, Brasil\\
\llap{$^{10}$} Department of Astrophysics, Radboud University Nijmegen, The Netherlands\\
\llap{$^{11}$} Institut f\"ur Prozessdatenverarbeitung und Elektronik, KIT, Germany\\
\llap{$^{12}$} Faculty of Natural Sciences and Engineering, Universit\"at Siegen, Germany\\
\llap{$^{13}$} Institute for Space Sciences, Bucharest-Magurele, Romania\\
\llap{$^{14}$} Fachbereich C, Physik, Universit\"at Wuppertal, Germany\\
\llap{$^{15}$} Department of Astrophysics, National Centre for Nuclear Research, {\L}\'{o}d\'{z}, Poland\\
\llap{$^{16}$} Department of Physics, University of Bucharest, Bucharest, Romania\\

E-mail: \email{frank.schroeder@kit.edu}       
}

\abstract{}

\FullConference{35th International Cosmic Ray Conference --- ICRC2017\\
		10--20 July, 2017\\
		Bexco, Busan, Korea}

\begin{document}

\section*{Abstract}
LOPES was the radio extension of the KASCADE-Grande particle-detector array consisting of up to 30 antennas measuring the radio emission of cosmic-ray air showers between 40 and 80 MHz with an energy threshold of around 100 PeV. 
Even with the external trigger by the particle detectors, the separation of the air-shower signal from the radio background was difficult in the noisy environment of the Karlsruhe Institute of Technology. 
For the typical event this was only possible because of the digital, interferometric beamforming technique pioneered by LOPES for cosmic-ray detection. 
Using this technique LOPES made important discoveries with respect to the radio emission of air showers and their relation to the shower properties, such as its energy and its longitudinal development. 
By now, practically all results have been confirmed by subsequent antenna arrays, but regarding digital interferometry the LOPES results are still unique. 
Lately we completed an end-to-end pipeline for CoREAS simulations of the radio emission including measured background and all known detector responses as well as the interferometric analysis technique.
As result we present an update on the reconstruction of the most important shower parameters: arrival direction, energy, and $X_\mathrm{max}$.

\section{Introduction}
LOPES was the first antenna array for air showers using computing-intensive, interferometric analysis techniques. 
This was key for the success of the experiment pioneering the digital era of the radio detection \cite{HuegeReview2016, SchroederReview2016}. 
When LOPES started operation as LOFAR prototype station in 2003 the radio emission by air showers was not yet fully understood. 
While the dominant geomagnetic nature of the emission was known \cite{Allan1971}, the weaker Askaryan emission was not yet experimentally confirmed for air showers and the refractive index of the air was mostly ignored.
Consequently, it was assumed that the radio signal is best visible below $100\,$MHz, which was one of the reasons why the band of $40-80\,$MHz was chosen for LOPES. 
Now it is known that the refractive index changes the coherence condition around the Cherenkov angle such that the geomagnetic and Askaryan signals extend up to several GHz \cite{CROME_PRL2014}, i.e., LOPES likely would have profited from a broader frequency band. 
Nevertheless and despite the high background due to external Galactic and anthropogenic noise, LOPES detected several 100 air showers.
It measured the arrival direction and energy of these air showers with an accuracy competitive with other techniques. 
LOPES also demonstrated the concept of two methods for the reconstruction of the position of the shower maximum, $X_\mathrm{max}$ \cite{2014ApelLOPES_wavefront, 2014ApelLOPES_MassComposition}, but was much less accurate than the later and denser LOFAR array \cite{LOFARNature2016}. 

The final LOPES analysis presented here is based on comparisons of the radio measurements with the coincident measurements by KASCADE-Grande and with end-to-end CoREAS simulations \cite{HuegeCoREAS_ARENA2012}. 
Already during the operation of LOPES regular comparisons with REAS simulations were essential to improve the simulations.
The predictions of the latest version of REAS, the CoREAS extension of the CORSIKA simulation code, are compatible with LOPES measurements on an absolute scale \cite{2015ApelLOPES_improvedCalibration}. 
Therefore, we use CoREAS to study the properties of the LOPES experiment in end-to-end simulations including all known detector effects. 
In particular, we have checked and updated the methods used for earlier results and compare predictions for expected reconstruction accuracies to the ones experimentally achieved. 
Meanwhile, all active LOPES members have dedicated themselves to newer experiments such as AERA \cite{AERAantennaPaper2012}, LOFAR \cite{LOFARNature2016}, and Tunka-Rex \cite{TunkaRex_NIM_2015}. 
Thus, the heritage of LOPES survives in these collaboration and has resulted in several common activities, e.g., the absolute calibration of LOFAR and Tunka-Rex with the LOPES calibration source \cite{NellesLOFAR_calibration2015}, and the subsequent comparison of the KASCADE and Tunka energy scales \cite{TunkaRexScale2016}.

\begin{figure}[p]
  \centering
  \includegraphics[width=0.99\linewidth]{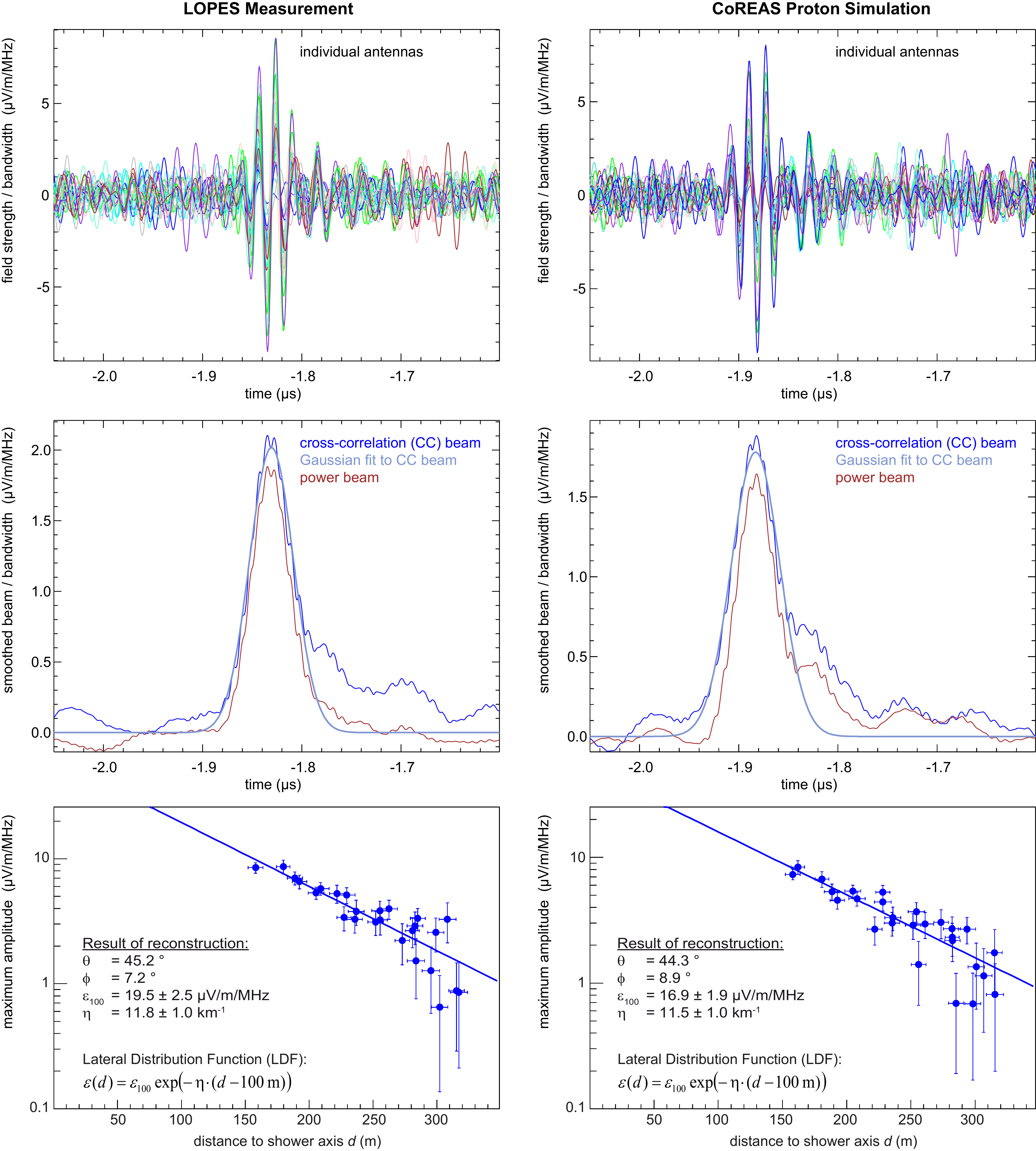}
  \caption{Example event triggered by KASCADE-Grande with a shower core outside the antenna array. 
  The KASCADE-Grande reconstruction used as true input for the simulations is: azimuth $\phi = 8.6^\circ$, zenith $\theta = 44.1^\circ$, energy $E = 3.4 \cdot 10^{17}\,$eV. 
  Thus, the event is far above the detection threshold and, unlike for many other events, the signal can clearly be seen by eye in individual antennas.
  \emph{Left:} LOPES measurement. 
  \emph{Right:} CoREAS end-to-end simulation applying all known detector effects. 
  \emph{Top:} Traces of individual antennas after a time shift corresponding to the arrival time of the radio wavefront in each antenna. 
  \emph{Middle:} Cross-correlation and power beam after smoothing by block-averaging over $37.5\,$ns. 
  \emph{Bottom:} Lateral distribution of the maximum amplitude measured in each antenna.}
  \label{fig_exampleEvent}
\end{figure}

\section{Experimental Setup, Measurements, and Simulations}
LOPES was an antenna array of about $200\,$m diameter in the KASCADE-Grande particle-detector array operating from 2003-2013.
To large parts LOPES was overlapping with the denser KASCADE particle-detector array providing accurate measurements of air showers \cite{Apel2010KASCADEGrande}. 
Starting with 10 antennas \cite{FalckeNature2005}, LOPES was soon extended to 30 antennas that at first were all aligned in east-west direction. 
Later, half of the antennas were rotated to north-south direction, and in 2009 the antennas were replaced by the LOPES-3D setup consisting of 10 tripole antennas \cite{Apel2012_LOPES3D}. 
All antennas were externally triggered which enabled a later correlation with the air-shower parameters reconstructed by KASCADE-Grande. 
Furthermore, 10 additional antennas operated in self-trigger mode (LOPES$^\mathrm{STAR}$), but recorded only a few events due to the high background. 
Thus, the development of self-trigger algorithms was continued by experiments at radio-quiet locations \cite{RAugerSelfTrigger2012, TREND2011}. 

The main results of LOPES are based on about 500 events with $E > 10^{17}\,$eV recorded with the east-west aligned antennas. 
In most of the events the radio signal cannot be distinguished from the background in individual antennas, but only by an interferometric combination of all antennas that is possible thanks to a nanosecond-precise time calibration \cite{SchroederTimeCalibration2010}. 
Applying the technique of cross-correlation beamforming, the individual traces are shifted in time according to the arrival time of the hyperbolic radio wavefront, i.e., the beamforming procedure implicitly reconstructs the arrival direction and the opening angle of the wavefront.
Once the radio signal is identified against the background, its amplitude can be determined in individual antennas to reconstruct the lateral distribution of the radio signal \cite{2010ApelLOPESlateral}. 
More details on this procedure are available in several publications, e.g., in the last PhD thesis related to LOPES \cite{LinkPhDThesis2016}. 
The amplitude of the cross-correlation beam as well as the amplitude of the lateral distribution at a specific reference distance are both good energy estimators \cite{2014ApelLOPES_MassComposition}, while the slope of the lateral distribution \cite{2014ApelLOPES_MassComposition, 2012ApelLOPES_MTD}, and the opening angle of the wavefront \cite{2014ApelLOPES_wavefront} can be used to estimate $X_\mathrm{max}$. 

The new and probably final LOPES results are based on a realistic set of end-to-end simulations. 
For each measured LOPES event two CoREAS simulations were produced, one for a proton, one for an iron nucleus as primary particle, using the shower axis and energy reconstructed by KASCADE-Grande as input \cite{2013ApelLOPESlateralComparison}.
Then, the detector responses of the antennas and the subsequent signal chain have been applied and real measured background has been added to the simulations. 
Subsequently, the simulations were analyzed in the same way as measured data including the cross-correlation beamforming procedure (see figure \ref{fig_exampleEvent} for an example event). 
Nonetheless, these end-to-end simulations fall still short of some experimental uncertainties, such as the uncertainty of the shower core. 
Hence, this procedure gives a realistic expectation of the achievable precision by LOPES for quantities whose uncertainties are dominated by background, and a best-case scenario for quantities dominated by systematic uncertainties.
Comparing the results of the end-to-end simulations to the LOPES measurements, thus, also provides indication in which aspects LOPES was limited by the high background of the site and where the performance was limited by other effects.

\begin{figure}[t]
  \centering
  \includegraphics[width=0.49\linewidth]{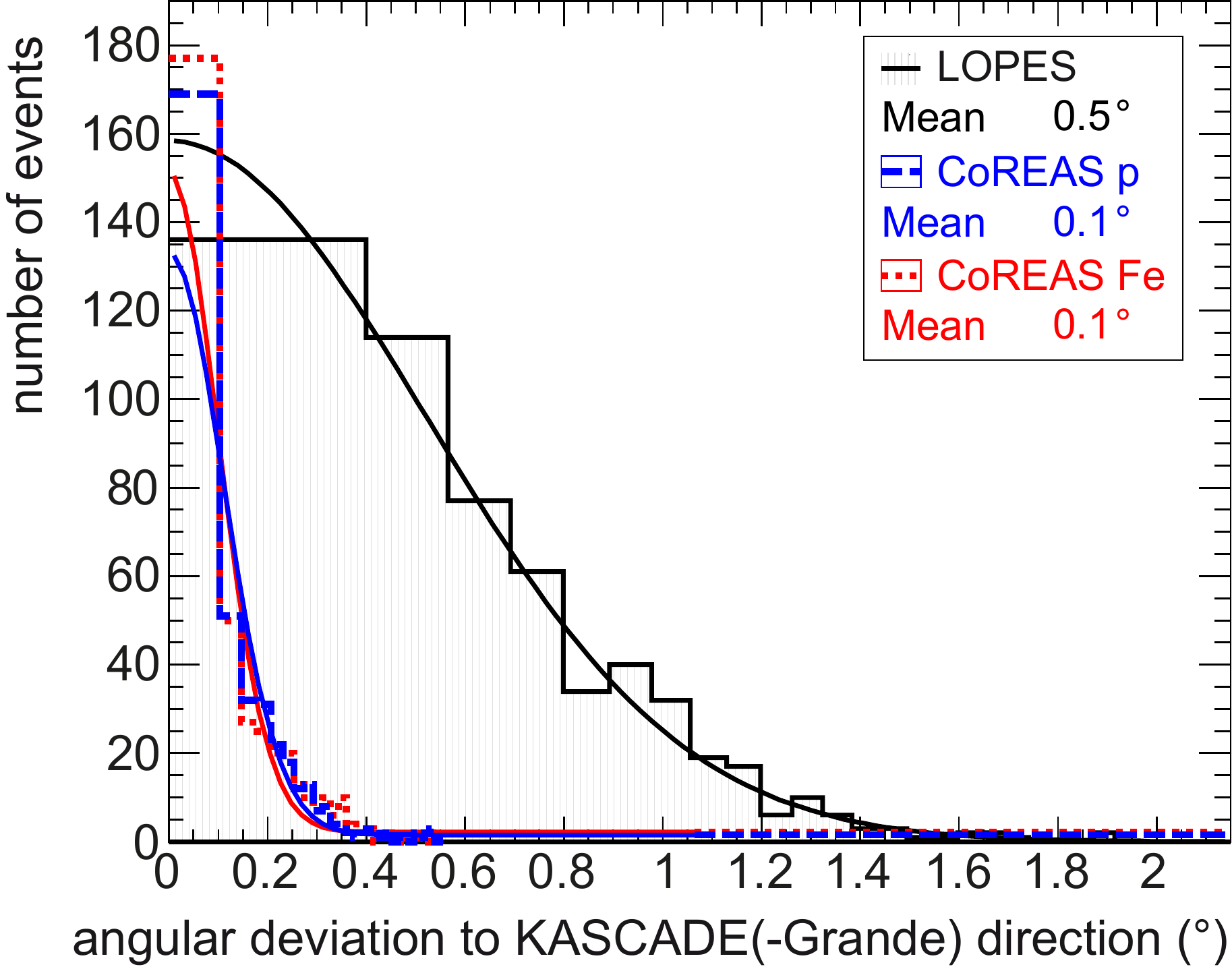}

  \caption{Difference between the true and reconstructed arrival directions for the simulations, and between KASCADE(-Grande) and LOPES for the measurements. 
  Simulations without noise (not shown) have a mean deviation of only $0.05^\circ$. 
  Each bin covers an equal solid angle.}
  \label{fig_angularDeviation}
\end{figure}

\section{Results}

\textbf{Arrival Direction: }
The arrival direction is reconstructed by maximizing the amplitude of the cross-correlation beam. 
Its accuracy (including bias and precision) is estimated by comparing the LOPES and KASCADE-Grande measurements. 
While earlier LOPES analyses based on a spherical instead of a hyperbolic wavefront achieved a direction accuracy of about $1^\circ$ \cite{NiglDirection2008}, the latest version of our standard analysis pipeline achieves about $0.5^\circ$ (figure \ref{fig_angularDeviation}). 
This value is only an upper limit on the LOPES accuracy, since it might be dominated by the resolution of KASCADE(-Grande) that is not known accurately for the used set of events. 
The end-to-end simulations reveal that even with realistic background a direction accuracy of $0.1^\circ$ should be possible.
Anyway, due to the magnetic deflection of charged cosmic rays on their way to Earth a further improvement of the angular resolution is uncritical, unless any neutral particles were detected in the energy range above $10^{17}\,$eV.

\begin{figure*}[t]
  \centering
  \includegraphics[width=0.4\linewidth]{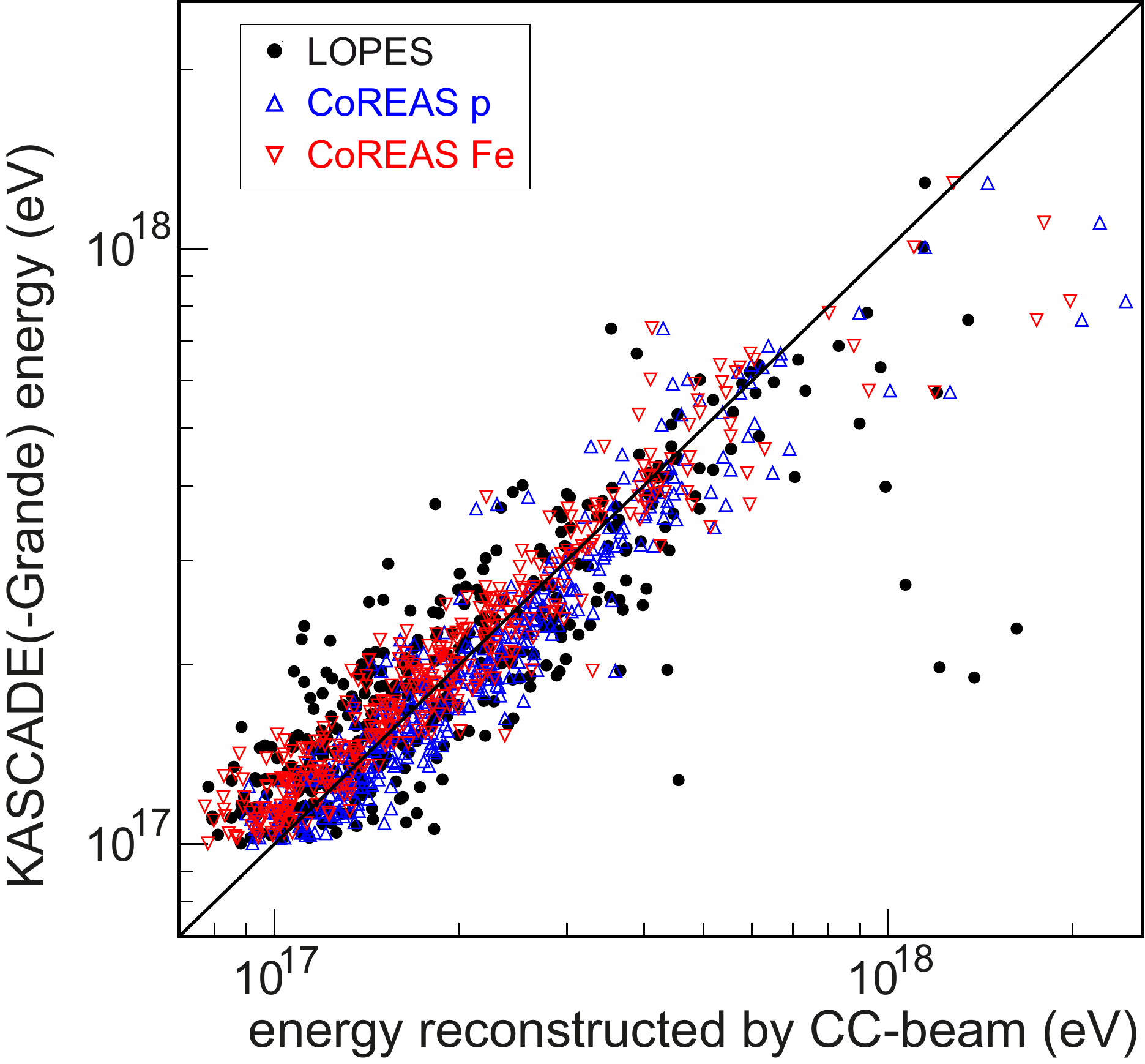}
  \hfill
  \includegraphics[width=0.49\linewidth]{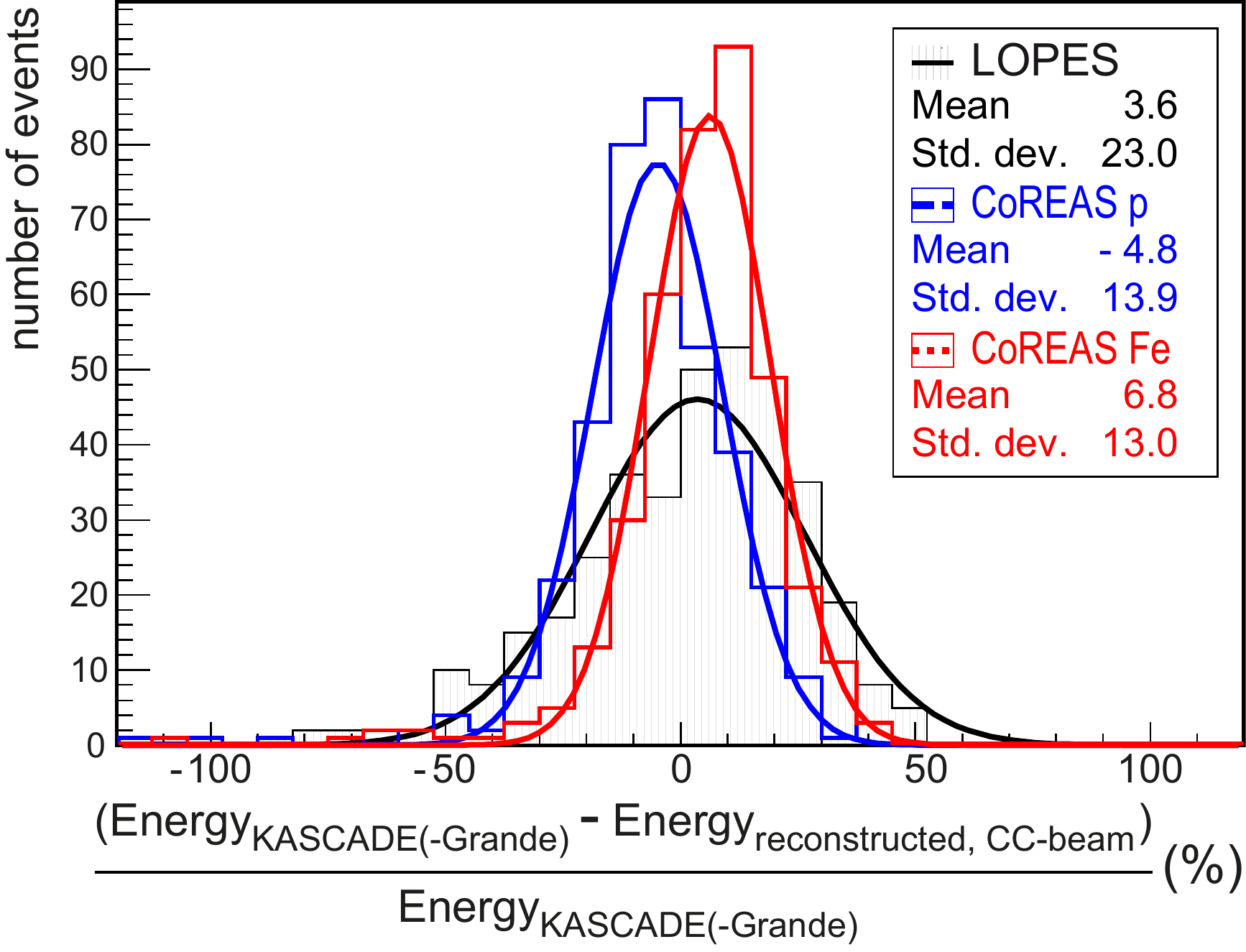}
  \caption{Comparison of the energy reconstructed by cross-correlation beamforming (LOPES measurements / CoREAS end-to-end simulations including background) to the KASCADE-Grande energy used as 'true' input for the simulations.
  The outliers are partly due to a small east-west component of the geomagnetic Lorentz vector, but the reason for some outliers is still unknown and not due to thunderstorms \cite{2011ApelLOPES_Thunderstorm}.
  }
  \label{fig_CCbeamEnergy}
\end{figure*}

\textbf{Energy: }
The energy of the air shower is reconstructed by LOPES in two ways: 
first, by the amplitude of the cross-correlation beam after correction for the mean distance of the antennas to the shower axis; 
second, by the amplitude at a reference distance of the order of $60-100\,$m, where the optimum distance depends on the zenith angle \cite{2014ApelLOPES_MassComposition}. 
Both methods yield a similar precision for the energy, and the second method has been studied in detail already with measurements and simulations.
With the end-to-end simulations we have now studied again the first method confirming the essence of earlier results and revising some details \cite{HornefferIcrc2007}.

In the new analysis we normalize the amplitude of the cross-correlation beam ($CC$) not anymore by $(1-\cos \alpha)$ to the geomagnetic angle $\alpha$, but instead by the east-west component of the unit vector of the Lorentz force $P_\mathrm{EW} = (\hat{v} \times \hat{B})$, which is orthogonal to the shower axis $v$ and to the geomagnetic field $B$ (the same correction as applied in \cite{2014ApelLOPES_MassComposition}). 
We keep the exponential correction for the mean distance to the shower axis $d_\mathrm{mean}$, since more complicated corrections, e.g., a correction corresponding to a Gaussian lateral distribution function, did not result in significantly higher precision of the reconstructed energy. 
Moreover, we further simplify the reconstruction by removing any correction for the zenith angle, since we find no significant dependence of the cross-correlation amplitude within the used range of $\theta < 45^\circ$.
Thus, the shower energy $E$ is reconstructed as:

\begin{equation}
 E = k \cdot \frac{CC}{P_\mathrm{EW}\cdot\exp{(-d_\mathrm{mean}/180\,\mathrm{m})}}
\end{equation}

The proportionality constant $k$ used here is the average $k$ of proton and iron simulations in \cite{LinkPhDThesis2016}, i.e., $k = 41.9\,$PeV/(\textmu V/m/MHz) for 'KASCADE' events and $k = 36.7\,$PeV/(\textmu V/m/MHz) for 'Grande' events whose shower core is outside of the KASCADE and LOPES arrays. 
To which extent this difference is due to an insufficient distance correction and to which extent due to any other systematic effects requires deeper investigations at larger arrays. 
The simulations indicate that an energy precision of better than $15\,\%$ is possible when using the LOPES beamforming technique at realistic background (figure \ref{fig_CCbeamEnergy}). 
This is consistent with the larger spread observed for the measurements, since the energy resolution of KASCADE-Grande is about $20\,\%$. 

\begin{figure*}[t]
  \centering
  \includegraphics[width=0.75\linewidth]{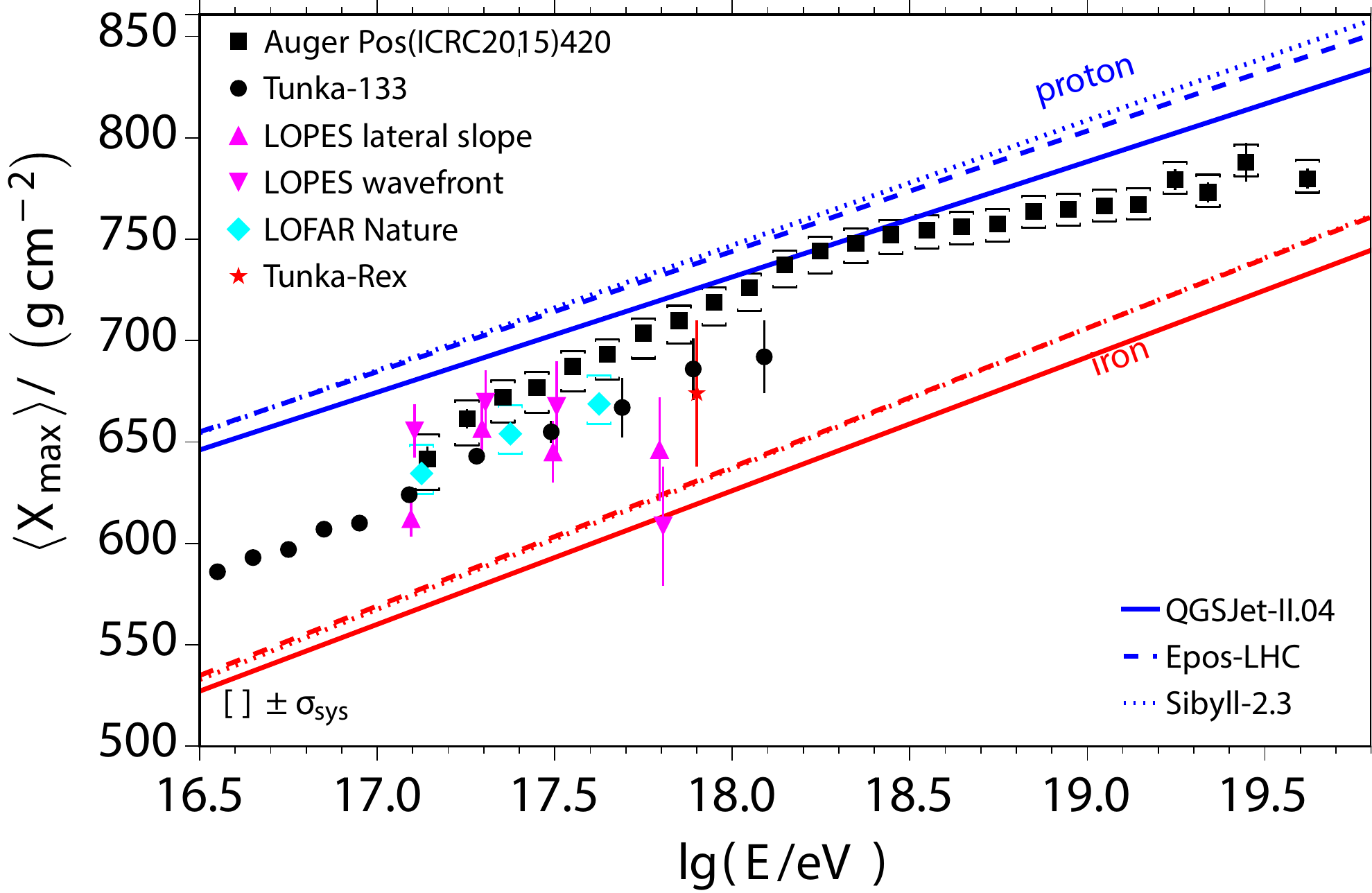}
  \caption{LOPES mean $X_\mathrm{max}$ in comparison to other radio arrays (color) \cite{LOFARNature2016, TunkaRex_Xmax2016}, or optical detectors (black) \cite{Tunka133_ISVHECRI2014, AugerHEATXmaxICRC2015}, and air-shower simulations \cite{RiehnPhDThesis2015}. 
  For LOPES the systematic uncertainties, e.g. due to selection biases, are not known (plot from Ref.~\cite{SchroederReview2016} with updated values for LOPES wavefront analysis).}
  \label{fig_XmaxOverview}
\end{figure*}

\textbf{Shower maximum: }
We used two methods to determine the position of the shower maximum. 
First, the slope of the lateral distribution: 
Since the slope changes only marginally in the end-to-end simulations compared to pure CoREAS simulations used earlier, we have not re-analyzed our $X_\mathrm{max}$ results based on this method \cite{2014ApelLOPES_MassComposition}. 
Second, the cone angle $\rho$ of the radio wavefront that for LOPES measurements is determined during cross-correlation beamforming: 
For the CoREAS simulations we had earlier determined $\rho$ based on the pulse arrival time in individual antennas \cite{2014ApelLOPES_wavefront}. 
Now we have studied this method applying exactly the same beamforming procedure on the end-to-end simulations as for the measured data. 
This leads to small changes in the reconstruction formula for $X_\mathrm{max}$, in particular a slightly weaker correction for the zenith angle $\theta$ \cite{LinkPhDThesis2016}:

\begin{equation}
 X_\mathrm{max} = 26143\,\mathrm{g/cm}^2 \cdot \rho \cdot \cos^{-1.24}\theta
\end{equation}
where $\rho$ (in rad) is the angle between the asymptotic cone of the hyperbolic wavefront and the plane perpendicular to the shower axis.
While the pure CoREAS simulations indicate that under ideal conditions an accuracy of better than $30\,$g/cm$^2$ would be possible when measuring arrival times in individual antennas, the end-to-end simulations yield a resolution of about $45\,$g/cm$^2$ without noise and almost $90\,$g/cm$^2$ with background. 
However, the precision of the LOPES measurements is even worse and estimated to about $130\,$g/cm$^2$ when using the wavefront method. 
The difference to the simulations indicates that in addition to background there are other relevant uncertainties, e.g., uncertainties of the shower direction and core. 
Despite of the large uncertainties both methods for $X_\mathrm{max}$ have been shown to work in principle with consistent results (see figure \ref{fig_XmaxOverview}).

\section{Conclusion}
By applying all known detector responses as well as measured background to CoREAS simulations of the radio emission for air showers, we could finally study the effect of background on the interferometric analysis method used by LOPES. 
The simulations show that the direction can be reconstructed by cross-correlation beamforming as accurately and precisely as $0.1^\circ$ and the energy with a precision of better than $15\,\%$.
This is consistent with the experimentally determined upper limits on the precision of $< 0.5^\circ$ and $< 23\,\%$ containing both the uncertainty of LOPES and KASCADE-Grande. 
Additionally the accuracy of the energy is limited by a $20\,\%$ scale uncertainty due to the absolute calibration of the antennas \cite{2015ApelLOPES_improvedCalibration}.
Moreover, the dependence of the cross-correlation beam on the steepness of the hyperbolic radio wavefront can be used to determine $X_\mathrm{max}$. 
However, for LOPES the uncertainties are even larger than for the reconstruction of $X_\mathrm{max}$ via the slope of the lateral distribution, and in both cases the uncertainties are too large for an interpretation in terms of mass composition. 
Nevertheless, these methods should be tested again either at very dense arrays such as the SKA \cite{HuegeSKA_ICRC2015}, or at large arrays for inclined showers, such as AERA \cite{AERAinclined_ARENA2016} or GRAND \cite{GRAND_ICRC2017}.


\bibliographystyle{JHEP}
\bibliography{icrc2017}

\end{document}